# A testable conventional hypothesis for the DAMA-LIBRA annual modulation

David Nygren
Physics Division, Lawrence Berkeley National Laboratory
1 Cyclotron Road, Berkeley, CA 94720
email: drnygren@lbl.gov

**Abstract:** The annual modulation signal observed by the DAMA-LIBRA Collaboration may plausibly be explained as a consequence of energy deposited in the NaI(Tl) crystals by cosmic ray muons penetrating the detector. Delayed pulses in the approximate energy range of interest have been observed as a sequel to energy deposited by UV irradiation. The same behavior may be reasonably expected to occur for energy deposited by any source of ionization or excitation. D-L can test this hypothesis by searching in current data for time correlations between muon events and pulses in the modulation energy range, or by renewed operation of the array at a sufficiently low temperature that freezes out the phenomenon.

**Introduction**

In 2008, the DAMA-LIBRA collaboration (D-L) published results claiming detection of WIMP Dark Matter, based on the observation of an annual modulation signal of greater than 8 σ significance [1]. Within the energy interval 2 – 5 $keV_{ee}$, the measured modulation is 0.0176 ±0.0020; results for slightly wider (2 – 6 $keV_{ee}$) or narrower (2 – 4 $keV_{ee}$) intervals are similar. [1] The modulation signal phase matches well the expected annual phase for standard assumptions about WIMP winds. The annual modulation signal observed by D-L is the result of a carefully executed ultra-low background experiment and is not generally challenged. Subsequent data taken in 2009 with detector updates is consistent with [1], and has slightly increased the statistical significance of the modulation signal [2]. More recent analysis of D-L is found in [3]. Altogether, D-L impose six WIMP criteria: a good cosine fit (1) with a period of one year (2) and phase near 2 June (3), modulation less than 7% (4), present only in a low-energy range (5), and in single-hit events only (6).

However, considerable controversy has followed the D-L interpretation of the modulation as evidence for WIMPs. The D-L claim narrowly evaded other results initially, but subsequent limits announced by CDMS, Xenon100, and CoGeNT have raised the level of tension between the D-L claim and combined limits [4 - 6]. A recent analysis including D-L, CoGeNT, and astrophysical data from Fermi-GLAST finds consistency for a WIMP mass of 7 – 8 GeV [7]. Nevertheless, the claim of the D-L modulation as evidence for WIMPs remains to be confirmed. Lacking confirmation, an obligation endures to consider any plausible explanation based on conventional physics. I propose here the existence of delayed pulses in the few $keV_{ee}$ range as a plausible scenario, based primarily on the complex and still incompletely understood response of alkali halide crystals to excitation and ionization. If this phenomenon is demonstrably present, a new kind of cosmogenic activation must be considered for experiments utilizing NaI(Tl) crystals.

---

[1] As usual, $keV_{ee}$ refers to a visible energy deposit equivalent to that of electrons.

**The detector, the observed modulation signal, and the rate**

The D-L detector at the LNGS is composed of 250 kg of NaI(Tl) supplied by St. Gobain, arranged in a 5 x 5 array of rectangular NaI bars, each 10.2 x 10.2 x 25.4 cm$^3$. Each bar is encapsulated in a radio-pure OFHC copper housing and viewed through special quartz light-guides at both end-faces by low-background PMTs. The sensitivities of the bar-PMT assemblies is excellent, and ranges from 5.5 – 7.5 photoelectrons (pe)/keV$_{ee}$. Software triggers require single-hit events (visible energy in one bar only), with pe time structure characteristic of scintillation (hundreds of ns) present in both PMTs, while rejecting short pulses (tens of ns) characteristic of PMT noise.

The rate data, seen here as figure 1 with their caption from [1], shows a nearly flat spectrum above 1.5 KeV$_{ee}$ up to 10 keV$_{ee}$ (suppressed zero). The shallow bump at ~3 keV$_{ee}$ is explained as a companion x-ray to a 1461 keV gamma-ray transition from a radioactive potassium impurity, and would therefore have no connection to the observed modulation signal. The sharp rise below 1.5 keV$_{ee}$ includes PMT noise, although these pulses include from ~3 – 12 pe in two PMTs and must satisfy all trigger conditions. As signal/background discrimination is less efficient below 2 keV$_{ee}$, a threshold of 2 keV$_{ee}$ was set by D-L for analysis. The rate is approximately 1 count/keV$_{ee}$/kg/day.

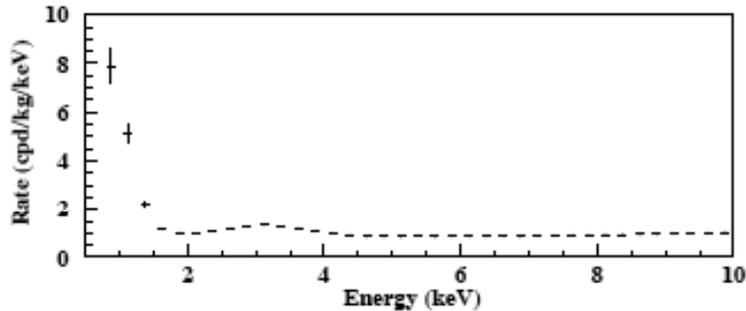

Figure 1: Cumulative low-energy distribution of the *single-hit* scintillation events (that is each detector has all the others as veto), as measured by the DAMA/LIBRA detectors in an exposure of 0.53 ton × yr. The energy threshold of the experiment is 2 keV and corrections for efficiencies are already applied.

The D-L modulation result [1] is reproduced here, as figure 2. The modulation is seen significantly only in a narrow energy band, 2 – 6 keV$_{ee}$, just above their chosen analysis threshold. This is a striking feature, given the nearly constant flux up to 10 keV$_{ee}$. The maximum amplitude of ~0.027 ±0.007 counts/ keV$_{ee}$/kg/day is found to occur at ~2.5 keV$_{ee}$. The average modulation over the 2 – 5 keV$_{ee}$ region is 0.0176 ±0.002 counts/keV-kg-day [1]. The modulation fades away with increasing energy, suggesting that, as signal declines, some background process rises coincidentally to maintain the same overall rate up to at least 10 keV$_{ee}$, a remarkable feature without obvious explanation. This character of the D-L data presents a challenge to explain, for any scenario, including WIMPs.

**Cosmic muons and neutrinos**

Seasonal temperature/density variations in the atmosphere lead to measured differences for in-flight meson decay rates within cosmic ray showers. The flux of cosmic muons penetrating the Gran Sasso National laboratory (LNGS) is thus a potential source of annual modulation. The seasonal variations observed in [8] for the LNGS do not follow strictly a simple cosine form, but on average have approximately the right phase relative to the least complicated assumptions for WIMP wind direction. Could the signal be due to low-energy atmospheric neutrinos associated with decays of pion and kaon parents of the muon flux? The answer is, of course, No, as this would require a neutrino-nucleus cross-section in excess of one millibarn.

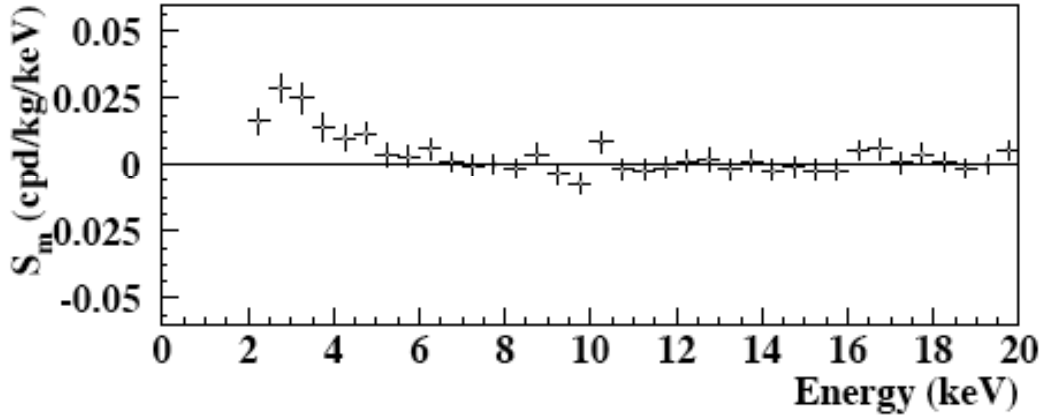

Figure 2. The D-L energy distribution of the combined total modulation signal for single-hit events is shown, again in keV$_{ee}$.

D-L discuss neutrons that cosmic muons make as they pass nearby the NaI(Tl) array. Relative to the observed modulation signal, the flux of these cosmic muon high-energy neutrons is calculated by D-L to fall short by two to three orders of magnitude, and D-L conclude that high energy neutrons from cosmic muons cannot be the source of their modulation signal. D-L attribute the multi-hit event class to high-energy neutrons, and no annual modulation is found in the multi-hit event class (not shown by D-L).

Yet, the neutron flux in Gran Sasso does display a seasonal variation, as discussed in [9]. Also in [9], neutron-nuclear interactions near thermal energies are considered in some detail. In this epithermal energy range, extremely complex resonant behavior is observed in neutron-nucleus interactions, argued comprehensively in [9] to be under-appreciated. If epithermal neutrons retain the annual modulation of their cosmic muon parents, which is to be expected, and are indeed the source of D-L signal, how can the multi-hit neutron-induced events not show modulation? Could these events arise predominantly from radioactivity, such as ($\alpha$,n) reactions within the detector? Multi-hit events induced by gamma-rays from radioactivity would also not show modulation, but should display recognizably different space-time correlations within the array. In any case, my goal here is not to argue that neutrons did, or did not, cause the annual modulation, but rather to introduce another plausible conventional background process that would, if present, mimic the D-L signal without neutron intermediaries.

Whenever a muon passes through one or more crystals in the D-L array of NaI, a gigantic signal occurs, vastly above the few keV$_{ee}$ region where the modulation is observed. Such huge signals are easily rejected. The muon flux would thus appear to have no direct connection to the annual modulation. I advance the hypothesis that muons interacting with the NaI(Tl) crystals induce *indirectly* the observed annual modulation signal by a mechanism leading to small delayed pulses.

## Radiation Damage & Phosphorescence in NaI(Tl)

While much of deposited energy within an alkali halide scintillator emerges in microseconds as the familiar scintillation, a substantial fraction of the remnant energy appears on a much longer time scale as a kind of phosphorescence, on time scales of minutes, hours, or even days. The appearance of delayed pulses corresponding to a few keV$_{ee}$ is contrary to intuition, but it turns out that this is indeed plausible. The basis for such a speculative hypothesis comes directly from St. Gobain, supplier of scintillation materials and products, including the NaI for D-L. St. Gobain Technical Information Note, #527, refers to "Effects of Ultraviolet (UV) Light on NaI, CsI, and BGO Crystals" [10].

Quoting from [10], "*__With mild exposure__ several pulses/second can be seen in the 6-10 keV region of a spectrum.*[2] If the crystal is stored in a dark area, this mild UV exposure will eventually disappear, although it may take from several hours to several days for the effects to stop." The note goes on to describe the impacts of greater exposure: "Severe exposure to UV will appear as a severe light leak to the PMT with overall loss in Pulse Height and Pulse Height Resolution. At this point no visible color centers can be seen, but effects to the NaI can be irreparable." Note #527 further adds that damage from extreme cases of UV exposure leads ultimately to a muddy brown discoloration.

Intuitively, it seems extremely unlikely that the observed pulses referred to in note #527 could be pile-up of random single-photons of phosphorescence: 6 -10 keV pulses imply ~30 - 50 photoelectrons within a few microseconds. Although Note 527 refers to 6 – 10 keV, it seems possible that only D-L has a setup with backgrounds low enough to explore the energy region below 6 keV$_{ee}$. Could D-L be seeing pulses in the few keV$_{ee}$ range as a kind of phosphorescent cascade – a release of energy on minutes-to-days time-scales that would mask their origin by muons? As far as [1] reports, D-L do not consider this possibility.

*Do muons deposit enough energy in the NaI(Tl) array to support such latent processes?*

According to D-L [1], the overburden of the Gran Sasso Laboratory reduces the high-energy muon flux to about 20 muons/m$^2$/day. The effective area of the D-L array is on the order of 1/4 - 1/3 m$^2$, suggesting that about five or six muons/day hit the array. Taking the NaI(Tl) density of 3.7 g/cm$^3$ and a muonic dE/dx of ~2 MeV/g/cm$^2$, a typical muon is found to deposit perhaps about 350 MeV in the array. The total muonic energy input is then about ~2 GeV/day.

The D-L spectrum is roughly 1.1 ± 0.1 count/keV$_{ee}$/kg/day (with ~1.7% annual modulation in the 2 – 5 keV$_{ee}$ range). Taking the modulation in the range 2 – 5 keV$_{ee}$, then the muons must deposit at least 3 x 1 keV bins x 1.1 cts/keV-kg-day x 3.5 keV (average energy) x 250 kg = ~2900 keV/day. This is about 1.5 parts per thousand of the ~2 GeV/day deposited by muons. There is no energy crisis – the muons do provide sufficient energy. The question then becomes: where does the energy go?

---

[2] Emphasis added.

Five distinct avenues account for most of it.
1. *Scintillation* The St. Gobain Technical Note for NaI crystal characteristics states a yield of 38 scintillation photons per keV$_\gamma$ (which I shall presume is equal to the yield from muons). Thus it takes ~26 eV to make one scintillation photon in NaI(Tl), whereas the characteristic scintillation at 420 nm corresponds to ~3 eV. So scintillation (μs time scales) only accounts for ~12% of deposited energy.
2. *Phosphorescence* is a commonplace phenomenon among materials that scintillate. For NaI(Tl), a decay time of 0.15 s has been measured, with a contribution of ~ 9% relative to the faster scintillation component [11]. In addition, other much longer time constants have also been observed, on the order of several days [12].
3. *Radiation damage*, in the form of color centers and other relatively stable, complex defect clusters. There is an example of very severe damage (brown coloration) caused by prolonged exposure to GeV electromagnetic showers. However, most of this damage was annealed in a short time at temperatures of around 300° C [13].
4. *Ionization* None of this signal is collected directly. For most substances, about three times the band-gap energy is required to produce one free electron-hole pair. As the band-gap in NaI is about 5.8 eV, an expenditure of about 17 eV per electron-hole pair is suggested. It is argued in [14] that the luminous efficiency of NaI(Tl) is ~50%, *i.e.,* only half of the number of created electron/hole pairs result in the emission of photons. Taking the 26 eV/photon as a good measure, then only ~13 eV per electron-hole pair are needed.[3] I suspect that the missing electron/ion pairs eventually do emit light or contribute to defect creation.
5. *Heat*, from direct phonon generation, and from sub-excitation electrons ultimately dissipating their kinetic energy. I have no clear way to estimate this fraction but as it must represent only low-energy excitations that lie below the major scintillation path, the contribution appears relatively small.

Known pathways of scintillation plus phosphorescence thus account for only about 13% of energy deposited. Much of the remaining ~86% of energy is available in NaI(Tl) to fuel complex avenues of long-term energy storage, with possible long-term release through annealing. For D-L, this implies availability of perhaps more than 1,700,000 keV/day, whereas only ~2900 keV/day is needed to produce "signal pulses". The D-L signal requires only 0.15% of the available deposited energy.

*How does UV energy at the ~5 eV level absorbed individually by atoms in the lattice become collectively released as a light pulse equivalent to 1000's of eV?*

This question involves complex, incompletely understood solid-state physics. Note 527 alludes to this circumstance: "...the effects of UV light to NaI have never been closely studied..." The luminous yield and decay time in NaI(Tl) both decrease strongly with increasing temperature, indicating the presence of quenching pathways with activation energies near room temperature. Thermally induced transitions are well known in solid-state physics, and one class of dosimeters is a good example of this process. The activation energies associated with transitions can be determined, for example, by an Arrhenius plot.

---

[3] On the other hand, CsI(Tl) appears to be ~94% efficient; the delayed pulse scenario may not be a significant process in this material.

Impurity and lattice defect migration can occur rapidly in many crystalline materials such as silicon and germanium semiconductors. For the alkali halide crystal KI a migrating excitonic mechanism has been advanced by Hersh [15] to explain **both** luminescence and color center formation. Hersh argues that excitonic transitions occur at UV energies, below ionization, leading to essentially identical responses to UV (6 – 10 eV) and x-rays. In other words, both cause luminescence and color center formation identically. So it could be that, with excitonic migration, some color centers create aggregates of stress energy in the lattice, perhaps some of which initiate a later cascade of energy release (e.g. San Andreas fault). The slow aggregation of latent energy is the key idea to generate late pulses. This proposed process is supported only by St. Gobain note #527 [10] and is the most speculative aspect of this hypothesis. Non-proportionality, with up to 20 – 40% higher response in the region of 10 keV relative to MeV energies is observed in common alkali-halide scintillators [16,17]; while non-proportionality may be unrelated to the present hypothesis, the phenomenon further exemplifies the complexities present.

### *Does the amplitude of the muon seasonal variation match that observed by D-L?*

The annual modulation measured by LVD [8] is 0.015 ±0.0006. Macro, with fewer cycles measured a similar but less precise value [18]. The LVD amplitude is equal, within error, to the observed D-L modulation amplitude. The near equality implies, in this scenario, that essentially all the observed D-L modulation is due to muons. Muons must contribute [0.0176 cts/keV$_{ee}$-kg-day (the D-L modulation)]/[0.015 (the muon annual modulation)] = 1.3 ±0.2 cts/keV$_{ee}$-kg-day. This value is in good agreement with the observed level of the D-L spectrum from 2 – 10 keV$_{ee}$, about 1.2 cts/keV$_{ee}$-kg-day (with accommodation for the 3 keV potassium bump).

In this scenario, the modulation in D-L data is diluted by energy deposited in the crystals by other sources of background due to radioactive decays. While the spectrum shown by D-L only covers the 1 – 10 keV$_{ee}$ range, it is possible to estimate the contribution from potassium decays. The 3 keV bump contributes approximately 1.5 keV/kg/day. If 90% of the potassium decays are tagged by the 1461 keV gamma, then the total/day is ~200 MeV; with these assumptions, the dilution would be ~10%, consistent within the error.

### *Does the phase of the muon annual modulation match that observed by D-L?*

I argue here that the phase of D-L annual modulation signal and the seasonal muon flux variation phase have not been shown to be inconsistent. The LVD muon flux data [8] have much higher statistical precision than D-L and, providentially, cover the six years of D-L data. The LVD data do display significant departures from a simple cosine behavior, contributing to a $\chi^2$ of 577 for 362 degrees of freedom for the simple cosine fit. The LVD muonic phase maximum is found at July 5 ±15 days, whereas D-L [1,2, 19] quote a phase maximum on day 146 ±7 (May 26). D-L claim the discrepancy between July 5 and May 26 is 5.9 σ. Three problems exist regarding this claim. First, in their assertion that the D-L phase is 5.9 sigma from July 5, D-L inappropriately ignore the 15 day error given by LVD. On this basis alone, the discrepancy is less than 3 sigma. Second, the most recent phase result reported [3] is now day 152 ± 7 days. Third, the 7 day phase uncertainty quoted by D-L is obtained in a fit to the D-L modulation data *only* if the period of 365.4 days is fixed, essentially introducing a *prior*. If the period is not fixed, the phase error can be shown to increase by a large factor, on the order of three, due to correlation in fitting. Since the period is a central found result, it is inappropriate to quote an error on modulation phase with the period fixed *a priori,* even if the free fit gives a result closely matching one year. Taking realistic errors together, no truly significant discrepancy exists concerning phase at this point. Ideally, a fit for consistency of D-L modulation to the LVD muon seasonal variation should be done on an annual basis; has that fit has been done? Instead of a criterion that the muons must display a good cosine fit, the better test is whether the D-L signal reasonably matches the more statistically precise muon seasonal variation. At present, muons cannot be ruled out on the basis of phase inconsistency.

### The delayed pulse hypothesis

As argued above, it seems fairly certain that UV, x-rays, and muons all deposit stored energy similarly within the NaI lattice. And, with the insights suggested by St. Gobain Note #527 and other studies referenced, it does not seem implausible to propose that a small portion of this energy is returned as delayed scintillation pulses as a consequence of some poorly understood self-annealing process. This hypothesis can be tested.

### What to do?

Seven possibilities come to mind.

1. D-L could review their data to check for time and space correlation of counts in the 2 – 6 and 6 – 10 $keV_{ee}$ ranges with the passage of muons through the various crystals in the NaI array. The time scale for release of energy as phosphorescent pulses is unknown. However, the presence of a non-zero correlation on any time scale would imply that this mechanism is active, raising a red flag.

2. D-L could embark on a new set of annual cycles, running the setup unchanged except for lower operating temperature. The goal would be that, at an appropriately reduced temperature, the annealing time would become much longer than a year, washing out any muon-induced modulation. While this test would be time consuming, the effort would still be much less than building any new apparatus somewhere else, where the phase of seasonal variations of muons is known to be different. Prior to this effort, some of the following tests should be done to ensure a conclusive result.

3. New experimental data could be taken to explore this proposed mechanism of muon-induced phosphorescence. A single high-purity low-background NaI(Tl) scintillator with two low-radioactivity PMTs could be operated in a well-shielded underground laboratory less deep than LNGS. Ideally, the muon flux and seasonal variation should also be well measured at this location. The idea is to reproduce the D-L conditions to the extent possible, except with much higher muon flux. If the muonic phosphorescence process is indeed responsible, the "signal" will be much higher per kg than in D-L, but show the same relative modulation.

4. Alternatively, such a well-shielded detector could be exposed to a pulsed x-ray source to evoke phosphorescence. Results could be obtained at different temperatures. Based on the arguments above [15-17], it seems likely that the energy deposition and transfer processes in NaI(Tl) for UV, muons, and x-rays are equivalent, and thus could provide a meaningful test. Such a test would show the appropriate operating temperature for point 2.

5. A study of the activation energies in NaI(Tl) may show the temperatures at which this proposed phosphorescence mechanism is elicited, or, conversely, frozen out. This knowledge would enhance the confidence of interpretation of any new data taken at various temperatures by D-L or by new efforts with other detectors.

6. DM-Ice is a new project that will place radio-pure NaI(Tl) crystals in the deep ice at the south pole. At a temperature of -30° C, exciton mobility is greatly lowered. The formation rate of energy aggregation sites, as argued above, will be correspondingly impaired. Thermally induced cascades leading to the pulses of interest will likely be frozen out, despite the fact that no basic physics has changed. Thus DM-Ice will very likely not see any annual muonic modulation. However, if the modulation found by D-L is seen by DM-Ice with the same phase and amplitude, an important step forward will have been made.

7. The presence of the 3 keV x-ray in the data is a prominent feature in the energy band of interest and introduces a complicating factor. A study of the D-L events where both 3 keV x-ray and 1461 keV gamma ray from potassium decays are detected should be undertaken to search for the presence of a significant annual modulation in this sub-class of events where no modulation is expected under conventional assumptions about radioactive decay.

## Summary

From any source of radiation, from UV, x-rays, neutrons, to muons, NaI(Tl) crystals display complex time behaviors in scintillation/phosphorescence. Strong temperature dependences exist in response, decay time (up to several days), and in annealing. Muons that strike the D-L array deposit ~1000 times more energy than is contained in the observed modulation signal. Phosphorescent pulses can plausibly explain the observed modulation signal in the 2 – 6 keV$_{ee}$ range as a consequence of muonic seasonal variation. The discrepancy in phases appears not to be significant, despite assertions by D-L. It is challenging to find any explanation, including WIMPs, for the absence of modulation above 5 keV$_{ee}$ while preserving the observed flat spectrum. The existing D-L data might show a crystal-by-crystal correlation between muon arrival time and pulses in the 2 – 5 keV$_{ee}$ range, which would provide an important hint. However, the annealing time in NaI(Tl) could be much longer than the mean time between muons (few/day) and still contribute strongly to an annual modulation. D-L could choose to run for several annual cycles at a much lower temperature, sufficiently low that the annealing time can be confidently taken as long enough to freeze in or wash out the muonic annual modulation. DM-Ice will likely, within this hypothesis, be too cold to see an annual modulation due to muons.

## Conclusions

At this point, the muon-induced delayed phosphorescence hypothesis, although speculative in several respects, plausibly satisfies the six WIMP dark matter requirements put forth in [1,2]. Several straightforward tests can be made. D-L should search for the presence in their data of a time correlation between the passage of muons through each NaI(Tl) crystal and pulses in the 2 – 6 and 6 – 20 keV$_{ee}$ ranges. Tests with a single radio-pure NaI(Tl) detector in a well-shielded underground laboratory could explore both muon- and x-ray source-induced phosphorescence as a function of temperature, and perhaps provide a robust determination whether the phenomenon exists with the relevant characteristics.

## Acknowledgments

I thank Bill Moses and Allesandro Bettini for several very useful comments.